# Novel Techniques to Derive Capacity Results for Multi-User Interference Channels

Reza K. Farsani, Amir K. Khandani
Department of Electrical and Computer Engineering
University of Waterloo, Waterloo, ON. Canada
Email: {r3khosra, khandani}@uwaterloo.ca

***Abstract:*** **The Interference Channels (ICs) represent fundamental building blocks of wireless communication networks. Despite considerable progress in network information theory, available capacity results for ICs, specifically those with more than two users, are still very limited. One of the main difficulties in the analysis of these networks is how to establish useful capacity outer bounds for them. In this paper, novel techniques requiring subtle *sequential application of Csiszar-Korner identity* are developed to establish efficient single-letter outer bounds on the sum-rate capacity of interference networks. By using the outer bounds, a full characterization of the sum-rate capacity is then derived for various multi-user ICs under specific conditions. Our capacity results hold for both discrete and Gaussian networks.**

## I. INTRODUCTION

A major challenge in developing wireless networks is to manage interference that is a resultant of simultaneous communication by different users. In network information theory, this problem is basically analyzed by the Interference Channel (IC), a scenario where some transmitters send independent messages to their respective receivers via a common media. Despite considerable progress in Shannon theory, the capacity region of ICs, even for the simple two-user case, is still open problem. A limiting expression for the capacity region in the general case was derived in [1]; however, computing the capacity through this expression is difficult.

Capacity bounds for the ICs have been studied in numerous papers. The two-user IC is considered in [1-12]. Specifically, a single-letter characterization of the capacity region has been derived for some special cases such as the IC with strong interference [2, 3], a subclass of deterministic ICs [4], a class of discrete degraded IC [5], and a subclass of one-sided IC [6]. In the concurrent works [7], [8] and [9], the authors established new capacity outer bounds for the two-user Gaussian IC. Using the derived outer bounds, they also identified a noisy interference regime for the channel. In this regime, treating interference as noise at each receiver is sum-rate optimal. Moreover, new sum-rate capacities were recently derived in [10] for the two-user *discrete* IC.

The multi-user ICs have been also studied widely in recent years. However, available capacity results are very limited. The so-called many-to-one and one-to-many ICs -special cases where interference is experienced, or is caused, by only one user- were considered in [13]. Capacity bounds were presented in [14] for a class of three-user deterministic ICs. An outer bound was presented in [15] for the three-user IC, and the sum-rate capacity was established for some certain Gaussian channels. A class of multi-user cyclic Gaussian IC was studied in [16] where the capacity region was approximately determined to within a constant gap; the authors also identified a strong interference regime for the Gaussian channel. Capacity region of a class of multi-user symmetric Gaussian ICs in the very strong interference regime was determined in [17] using lattice coding. In [18] a full characterization of the sum-rate capacity is provided for the degraded ICs. Finally, the strong interference regime initially derived in [2, 3] for the two-user IC is generalized to the multi-user ICs in [19].

One of the main difficulties in the analysis of multi-user ICs is how to establish useful capacity outer bounds for them. In this paper, we develop novel techniques requiring subtle *sequential application of Csiszar-Korner identity* [20] to establish efficient single-letter outer bounds on the sum-rate capacity of multi-user ICs. We demonstrate that our outer bounds can be applied to obtain the sum-rate capacity for various multi-user ICs under specific conditions. Our new capacity results hold for both discrete and Gaussian networks.

The rest of the paper is organized as follows. Preliminaries and definitions are given in Section II. The main results are presented in Section III.

# II. PRELIMINARIES

Throughout this paper, given a length-$n$ vector of random variables $X^n$, the notation $X^{n\backslash t}$ is defined as follows:

$$X^{n\backslash t} \triangleq (X^{t-1}, X_{t+1}^{n}), \qquad t = 1, \dots, n \tag{1}$$

The classical $K$-user interference channel is a communication scenario where $K$ transmitters send independent messages to their respective receivers via a common media. The channel is specified by $K$ input signals $X_i \in \mathcal{X}_i$, $K$ output signals $Y_i \in \mathcal{Y}_i$, $i = 1, \dots, K$, and the transition probability function $\mathbb{P}(y_1, y_2, \dots, y_K | x_1, x_2, \dots, x_K)$. Figure 1 depicts the channel model.

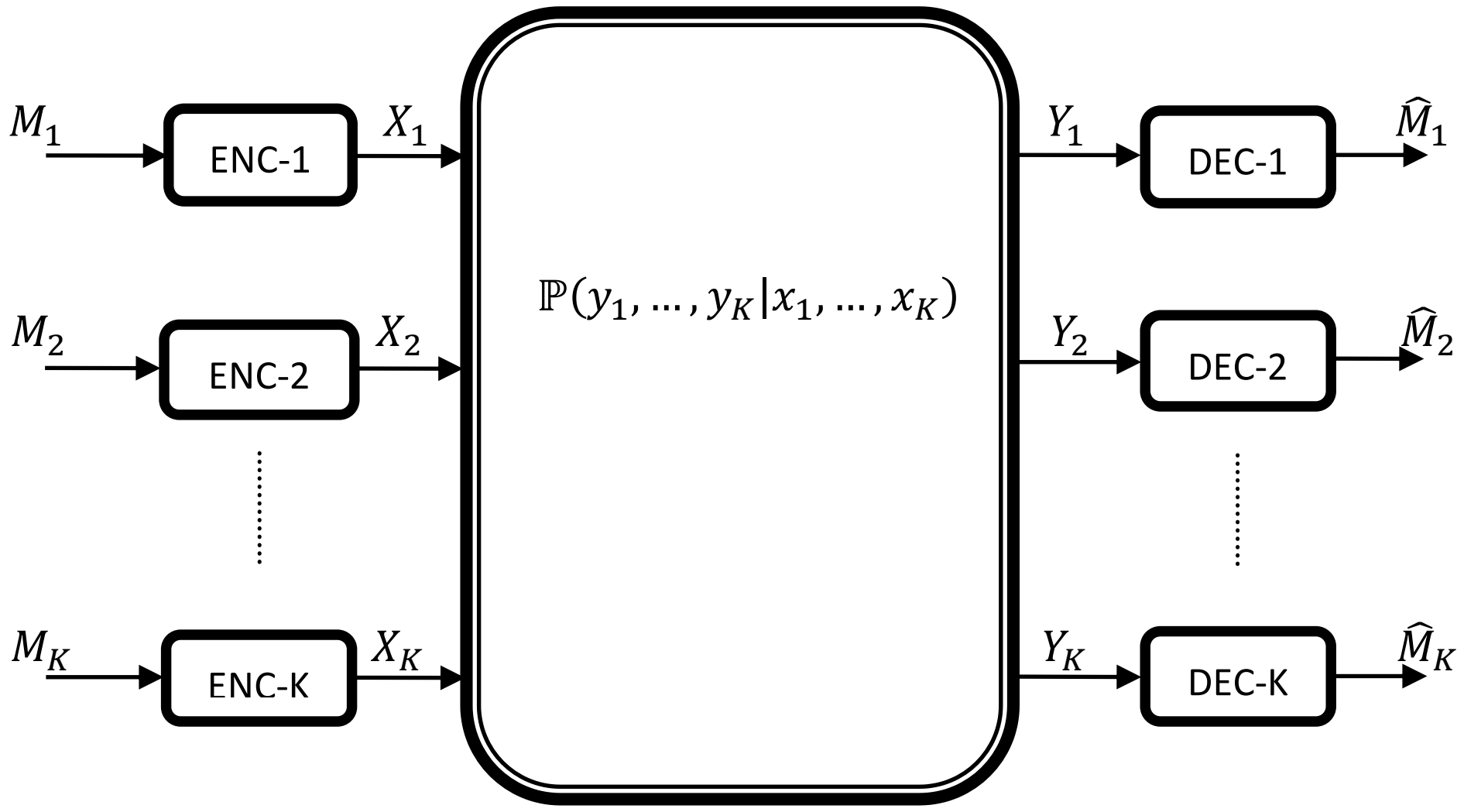

Figure 1. The K-user interference channel.

The Gaussian IC with real-valued input and output signals is given by:

$$\begin{bmatrix} Y_1 \\ \vdots \\ Y_K \end{bmatrix} = \begin{bmatrix} a_{11} & \cdots & a_{1K} \\ \vdots & \ddots & \vdots \\ a_{K1} & \cdots & a_{KK} \end{bmatrix} \begin{bmatrix} X_1 \\ \vdots \\ X_K \end{bmatrix} + \begin{bmatrix} Z_1 \\ \vdots \\ Z_K \end{bmatrix} \tag{2}$$

where the parameters $\{a_{ji}\}_{j,i=1,\dots,K}$ are (fixed) real-valued numbers, and the random variables $\{Z_j\}_{j=1}^{K}$ are zero-mean unit-variance Gaussian noises. Without loss of generality, one can assume that $a_{ii} = 1, i = 1, \dots, K$. The $i^{th}$ transmitter is subject to an average power constraint as: $\mathbb{E}[X_i^2] \leq P_i$, where $P_i \in \mathbb{R}_+$, $i = 1, \dots, K$.

In the following, we present some technical lemmas which have an essential role in our main derivations given in Section III. These lemmas are in fact proved in [21]. However, we have provided proofs of them in the Appendix for complicity.

Let $\mathcal{Y}_1, \mathcal{Y}_2, \mathcal{X}_1, \mathcal{X}_2, \dots, \mathcal{X}_{\mu_1}, \mathcal{X}_{\mu_1+1}, \dots, \mathcal{X}_{\mu_1+\mu_2}$ be arbitrary sets, where $\mu_1, \mu_2 \in \mathbb{N}$ are arbitrary natural numbers. Let also $\mathbb{P}(y_1, y_2 | x_1, x_2, \dots, x_{\mu_1}, x_{\mu_1+1}, \dots, x_{\mu_1+\mu_2})$ be a given conditional probability distribution defined on the set $\mathcal{Y}_1 \times \mathcal{Y}_2 \times \mathcal{X}_1 \times \mathcal{X}_2 \times \dots \times \mathcal{X}_{\mu_1} \times \mathcal{X}_{\mu_1+1} \times \dots \times \mathcal{X}_{\mu_1+\mu_2}$.

***Lemma 1.*** *Consider the inequality below:*

$$I(X_1, \dots, X_{\mu_1}; Y_1 | X_{\mu_1+1}, \dots, X_{\mu_1+\mu_2}) \leq I(X_1, \dots, X_{\mu_1}; Y_2 | X_{\mu_1+1}, \dots, X_{\mu_1+\mu_2}) \tag{3}$$

*If the inequality* (3) *holds for all PDFs* $P_{X_1 \dots X_{\mu_1} X_{\mu_1+1} \dots X_{\mu_1+\mu_2}}(x_1, \dots, x_{\mu_1}, x_{\mu_1+1}, \dots, x_{\mu_1+\mu_2})$ *with the following factorization:*

$$P_{X_1 \dots X_{\mu_1} X_{\mu_1+1} \dots X_{\mu_1+\mu_2}} = P_{X_1 \dots X_{\mu_1}}(x_1, \dots, x_{\mu_1}) P_{X_{\mu_1+1}}(x_{\mu_1+1}) P_{X_{\mu_1+2}}(x_{\mu_1+2}) \dots P_{X_{\mu_1+\mu_2}}(x_{\mu_1+\mu_2}) \tag{4}$$

*then, we have:*

$$I(X_1, \dots, X_{\mu_1}; Y_1 | X_{\mu_1+1}, \dots, X_{\mu_1+\mu_2}, D) \leq I(X_1, \dots, X_{\mu_1}; Y_2 | X_{\mu_1+1}, \dots, X_{\mu_1+\mu_2}, D) \tag{5}$$

*for all joint PDFs* $P_{DX_1 \dots X_{\mu_1} X_{\mu_1+1} \dots X_{\mu_1+\mu_2}}(d, x_1, \dots, x_{\mu_1}, x_{\mu_1+1}, \dots, x_{\mu_1+\mu_2})$ *where* $D \to X_1, \dots, X_{\mu_1}, X_{\mu_1+1}, \dots, X_{\mu_1+\mu_2} \to Y_1, Y_2$ *forms a Markov chain.*

*Proof of Lemma 1)* See Appendix A.■

**Corollary 1.** *For any* $\Omega \subseteq \{1, \dots, \mu_1\}$, *if the inequality (3) holds for all joint PDFs (4), then we have:*

$$I(\{X_1, \dots, X_{\mu_1}\} - \{X_i\}_{i\in\Omega}; Y_1 | \{X_i\}_{i\in\Omega}, X_{\mu_1+1}, \dots, X_{\mu_1+\mu_2}, D) \leq I(\{X_1, \dots, X_{\mu_1}\} - \{X_i\}_{i\in\Omega}; Y_2 | \{X_i\}_{i\in\Omega}, X_{\mu_1+1}, \dots, X_{\mu_1+\mu_2}, D) \tag{6}$$

*for all joint PDFs* $P_{DX_1 \dots X_{\mu_1} X_{\mu_1+1} \dots X_{\mu_1+\mu_2}}(d, x_1, \dots, x_{\mu_1}, x_{\mu_1+1}, \dots, x_{\mu_1+\mu_2})$ *where* $D \to X_1, \dots, X_{\mu_1}, X_{\mu_1+1}, \dots, X_{\mu_1+\mu_2} \to Y_1, Y_2$ *forms a Markov chain.*

***Lemma 2.*** *Consider the inequality below:*

$$I(U; Y_1 | X_{\mu_1+1}, \dots, X_{\mu_1+\mu_2}) \leq I(U; Y_2 | X_{\mu_1+1}, \dots, X_{\mu_1+\mu_2}) \tag{7}$$

*If the inequality* (7) *holds for all PDFs* $P_{UX_1 \dots X_{\mu_1} X_{\mu_1+1} \dots X_{\mu_1+\mu_2}}(x_1, \dots, x_{\mu_1}, x_{\mu_1+1}, \dots, x_{\mu_1+\mu_2})$ *with the following factorization:*

$$P_{UX_1 \dots X_{\mu_1} X_{\mu_1+1} \dots X_{\mu_1+\mu_2}} = P_{UX_1 \dots X_{\mu_1}}(u, x_1, \dots, x_{\mu_1}) P_{X_{\mu_1+1}}(x_{\mu_1+1}) P_{X_{\mu_1+2}}(x_{\mu_1+2}) \dots P_{X_{\mu_1+\mu_2}}(x_{\mu_1+\mu_2}) \tag{8}$$

*then, we have:*

$$I(U; Y_1 | X_{\mu_1+1}, \dots, X_{\mu_1+\mu_2}, D) \leq I(U; Y_2 | X_{\mu_1+1}, \dots, X_{\mu_1+\mu_2}, D) \tag{9}$$

*for all joint PDFs* $P_{DUX_1 \dots X_{\mu_1} X_{\mu_1+1} \dots X_{\mu_1+\mu_2}}(d, u, x_1, \dots, x_{\mu_1}, x_{\mu_1+1}, \dots, x_{\mu_1+\mu_2})$ *where* $D, U \to X_1, \dots, X_{\mu_1}, X_{\mu_1+1}, \dots, X_{\mu_1+\mu_2} \to Y_1, Y_2$ *form a Markov chain.*

*Proof of Lemma 2)* See Appendix B. ■

For the Gaussian networks, we need to the following variations of Lemmas 1 and 2. Let the outputs $Y_1$ and $Y_2$ be given as follows:

$$\begin{cases} Y_1 \triangleq a_1X_1 + a_2X_2 + \cdots + a_{\mu_1}X_{\mu_1} + a_{\mu_1+1}X_{\mu_1+1} + \cdots + a_{\mu_1+\mu_2}X_{\mu_1+\mu_2} + Z_1 \\ Y_2 \triangleq b_1X_1 + b_2X_2 + \cdots + b_{\mu_1}X_{\mu_1} + b_{\mu_1+1}X_{\mu_1+1} + \cdots + b_{\mu_1+\mu_2}X_{\mu_1+\mu_2} + Z_2 \end{cases} \tag{10}$$

where $Z_1$ and $Z_2$ are zero-mean unit-variance Gaussian random variables; also, $X_1, X_2, \dots, X_{\mu_1}, X_{\mu_1+1}, \dots, X_{\mu_1+\mu_2}$ are real-valued power-constrained random variables independent of $(Z_1, Z_2)$ and $a_1, a_2, \dots, a_{\mu_1}, a_{\mu_1+1}, \dots, a_{\mu_1+\mu_2}$ and $b_1, b_2, \dots, b_{\mu_1}, b_{\mu_1+1}, \dots, b_{\mu_1+\mu_2}$ are fixed real numbers. The goal is to determine sufficient conditions for this setup under which the inequality (5) (or (9)) holds for all joint PDFs $P_{DX_1\dots X_{\mu_1}X_{\mu_1+1}\dots X_{\mu_1+\mu_2}}(d, x_1, \dots, x_{\mu_1}, x_{\mu_1+1}, \dots, x_{\mu_1+\mu_2})$. The following lemmas present such conditions.

***Lemma 3.*** *Consider the Gaussian system in (10). If the following condition satisfies:*

$$\frac{a_1}{b_1} = \frac{a_2}{b_2} = \cdots = \frac{a_{\mu_1}}{b_{\mu_1}} = \alpha, \qquad |\alpha| \leq 1, \tag{11}$$

*then the inequality (5) holds for all joint PDFs* $P_{DX_1\dots X_{\mu_1}X_{\mu_1+1}\dots X_{\mu_1+\mu_2}}(d, x_1, \dots, x_{\mu_1}, x_{\mu_1+1}, \dots, x_{\mu_1+\mu_2})$ *where* $D$ *is independent of* $(Z_1, Z_2)$.

*Proof of Lemma 3)* See Appendix C.■

In fact, under the condition (11), given $X_{\mu_1+1}, \dots, X_{\mu_1+\mu_2}$, the signal $Y_1$ is a stochastically degraded version of $Y_2$. Also, we remark that the relation (11) is a sufficient condition under which (5) holds; however, in general the inequality (5) may not be equivalent to (11). It is essential to note that the condition (11) is not derived by evaluating (3) for Gaussian input distributions. Only for the case of $\mu_1 = 1$, the condition (11) can be equivalently derived by evaluating (3) for Gaussian input distributions.

***Lemma 4.*** *Consider the Gaussian system in (10). If (11) holds, then the inequality (9) is satisfied for all joint PDFs* $P_{DUX_1\dots X_{\mu_1}X_{\mu_1+1}\dots X_{\mu_1+\mu_2}}(d, u, x_1, \dots, x_{\mu_1}, x_{\mu_1+1}, \dots, x_{\mu_1+\mu_2})$ *where* $(D, U)$ *is independent of* $(Z_1, Z_2)$.

*Proof of Lemma 4)* The proof is in fact similar to Lemma 3. In essence, if the condition (11) holds, given $X_{\mu_1+1}, \dots, X_{\mu_1+\mu_2}$, the signal $Y_1$ is a stochastically degraded version of $Y_2$. Therefore, (9) is always satisfied for all joint PDFs $P_{DUX_1\dots X_{\mu_1}X_{\mu_1+1}\dots X_{\mu_1+\mu_2}}(d, u, x_1, \dots, x_{\mu_1}, x_{\mu_1+1}, \dots, x_{\mu_1+\mu_2})$. ■

Given these preliminaries, our main results are presented in the next section.

# IV. Main Results

In this section, first we establish a single-letter outer bound on the sum-rate capacity of the general multi-user ICs which satisfy certain less-noisy conditions. The proof of this outer bound includes a novel interesting technique requiring a sequential application of the Csiszar-Korner identity. We then obtain scenarios for which the derived outer bound is also achievable which yields the exact sum-rate capacity. Next, we present some generalizations of the ideas to derive other outer bounds which can be used to prove more capacity results.

***Theorem 1.*** *Consider the K-user IC shown in Fig. 1. Assume that the channel transition probability function satisfies the following conditions:*

$$I(U; Y_i|X_i, X_{i+1}, \dots, X_K) \leq I(U; Y_{i-1}|X_i, X_{i+1}, \dots, X_K), \qquad \textit{for all joint PDFs} \quad P_{UX_1X_2\dots X_{i-1}}P_{X_i}P_{X_{i+1}}\dots P_{X_K}, \quad i = 2, \dots, K \tag{12}$$

*The sum-rate capacity, denoted by* $\mathcal{C}_{sum}$, *is bounded as:*

$$\mathcal{C}_{sum} \leq \max_{P_Q P_{X_1|Q} P_{X_2|Q}\dots P_{X_K|Q}} \big( I(X_1; Y_1|X_2, X_3, \dots, X_K, Q) + I(X_2; Y_2|X_3, X_4, \dots, X_K, Q) + \cdots + I(X_K; Y_K|Q) \big) \tag{13}$$

*Proof of Theorem 1)* First note that, according to Lemma 2, the conditions (12) can be extended as follows:

$$I(U;Y_i|X_i,\dots,X_K,D)\le I(U;Y_{i-1}|X_i,\dots,X_K,D),\quad \text{for all joint PDFs}\quad P_{DUX_1X_2\dots X_K},\quad i=2,\dots,K \tag{14}$$

Now consider a code of length $n$ for the network with the rates $R_1,R_2,\dots,R_K$ for the users $X_1,X_2,\dots,X_K$, respectively. Using Fano's inequality, for $i=1,\dots,K$, we have:

$$\begin{aligned} nR_i &\le I(X_i^n;Y_i^n)+n\epsilon_{i,n}\\ &\overset{(a)}{\le} I(X_i^n;Y_i^n|X_{i+1}^n,\dots,X_K^n)+n\epsilon_{i,n}\end{aligned} \tag{15}$$

where $\epsilon_{i,n}\to 0$ as $n\to\infty$. Note that the inequality (a) in (15) holds because $X_i^n$ is independent of $X_{i+1}^n,\dots,X_K^n$. By adding the two sides of (15) for $i=1,\dots,K$, we obtain:

$$n\mathcal{C}_{sum}=n\sum_{i=1}^{K}R_i\le I(X_1^n;Y_1^n|X_2^n,\dots,X_{K-1}^n,X_K^n)+\dots+I(X_{K-1}^n;Y_{K-1}^n|X_K^n)+I(X_K^n;Y_K^n)+n\epsilon_n \tag{16}$$

where $\epsilon_n\to 0$ as $n\to\infty$. The gist of the proof is to provide a single-letter characterization of the expression on the right side of (16). We derive this by a sequential application of Csiszar-Korner identity [20] as well as by utilizing the conditions (14). Consider the last two mutual information functions in (16). We have:

$$\begin{cases} I(X_{K-1}^n;Y_{K-1}^n|X_K^n)=\sum_{t=1}^{n}I\left(X_{K-1}^n;Y_{K-1,t}\middle|X_K^n,Y_{K-1}^{t-1}\right)\\ I(X_K^n;Y_K^n)=\sum_{t=1}^{n}I\left(X_K^n;Y_{K,t}\middle|Y_{K,t+1}^n\right)\le\sum_{t=1}^{n}I\left(X_K^n,Y_{K,t+1}^n;Y_{K,t}\right)\end{cases} \tag{17}$$

Please precisely observe the style of applying the chain rule in each of the equations (81). Now, from (81) we derive:

$$\begin{aligned} &I(X_{K-1}^n;Y_{K-1}^n|X_K^n)+I(X_K^n;Y_K^n)\\ &\le\textstyle\sum_{t=1}^n I\left(X_{K-1}^n;Y_{K-1,t}\middle|X_K^n,Y_{K-1}^{t-1}\right)+\sum_{t=1}^n I\left(X_K^n,Y_{K,t+1}^n;Y_{K,t}\right)\\ &=\textstyle\sum_{t=1}^n I\left(X_{K-1}^n,Y_{K,t+1}^n;Y_{K-1,t}\middle|X_K^n,Y_{K-1}^{t-1}\right)+\sum_{t=1}^n I\left(X_K^n,Y_{K-1}^{t-1},Y_{K,t+1}^n;Y_{K,t}\right)\\ &\qquad\textstyle-\sum_{t=1}^n I\left(Y_{K,t+1}^n;Y_{K-1,t}\middle|X_{K-1}^n,X_K^n,Y_{K-1}^{t-1}\right)-\sum_{t=1}^n I\left(Y_{K-1}^{t-1};Y_{K,t}\middle|X_K^n,Y_{K,t+1}^n\right)\\ &=\textstyle\sum_{t=1}^n I\left(X_{K-1}^n;Y_{K-1,t}\middle|X_K^n,Y_{K-1}^{t-1},Y_{K,t+1}^n\right)+\sum_{t=1}^n I\left(Y_{K,t+1}^n;Y_{K-1,t}\middle|X_K^n,Y_{K-1}^{t-1}\right)\\ &\qquad\textstyle+\sum_{t=1}^n I\left(Y_{K-1}^{t-1},Y_{K,t+1}^n;Y_{K,t}\middle|X_K^n\right)+\sum_{t=1}^n I\left(X_K^n;Y_{K,t}\right)\\ &\qquad\textstyle-\sum_{t=1}^n I\left(Y_{K,t+1}^n;Y_{K-1,t}\middle|X_{K-1}^n,X_K^n,Y_{K-1}^{t-1}\right)-\sum_{t=1}^n I\left(Y_{K-1}^{t-1};Y_{K,t}\middle|X_K^n,Y_{K,t+1}^n\right)\\ &\overset{(a)}{=}\textstyle\sum_{t=1}^n I\left(X_{K-1}^n;Y_{K-1,t}\middle|X_K^n,Y_{K-1}^{t-1},Y_{K,t+1}^n\right)+\sum_{t=1}^n I\left(Y_{K-1}^{t-1},Y_{K,t+1}^n;Y_{K,t}\middle|X_K^n\right)\\ &\qquad\textstyle+\sum_{t=1}^n I\left(X_K^n;Y_{K,t}\right)-\sum_{t=1}^n I\left(Y_{K,t+1}^n;Y_{K-1,t}\middle|X_{K-1}^n,X_K^n,Y_{K-1}^{t-1}\right)\end{aligned} \tag{18}$$

where the equality (a) holds because, according to Csiszar-Korner identity, the $2^{nd}$ and the $6^{th}$ ensembles on the left hand side of (a) are equal. Now consider the second ensemble on right side of (a) in (18). We claim that:

$$\sum_{t=1}^{n}I\left(Y_{K-1}^{t-1},Y_{K,t+1}^n;Y_{K,t}\middle|X_K^n\right)\le\sum_{t=1}^{n}I\left(Y_{K-1}^{t-1},Y_{K,t+1}^n;Y_{K-1,t}\middle|X_K^n\right) \tag{19}$$

To verify the inequality (19), first note that we have:

$$\begin{cases} \sum_{t=1}^{n} I\left(Y_{K-1}^{t-1}, Y_{K,t+1}^{n}; Y_{K,t} \middle| X_K^n\right) = \sum_{t=1}^{n} I\left(Y_{K-1}^{t-1}, Y_{K,t+1}^{n}; Y_{K,t} \middle| X_{K,t}, X_K^{n\backslash t}\right) \\ \sum_{t=1}^{n} I\left(Y_{K-1}^{t-1}, Y_{K,t+1}^{n}; Y_{K-1,t} \middle| X_K^n\right) = \sum_{t=1}^{n} I\left(Y_{K-1}^{t-1}, Y_{K,t+1}^{n}; Y_{K-1,t} \middle| X_{K,t}, X_K^{n\backslash t}\right) \end{cases} \tag{20}$$

Considering (20), the inequality (19) is derived from the condition (14) for $i = K, U \equiv \left(Y_{K-1}^{t-1}, Y_{K,t+1}^{n}\right)$ and $D \equiv X_K^{n\backslash t}$. Now, by substituting (19) into (18), we obtain:

$$\begin{aligned} &I(X_{K-1}^n; Y_{K-1}^n | X_K^n) + I(X_K^n; Y_K^n) \\ &\quad \le \sum_{t=1}^{n} I\left(X_{K-1}^n; Y_{K-1,t} \middle| X_K^n, Y_{K-1}^{t-1}, Y_{K,t+1}^{n}\right) + \sum_{t=1}^{n} I\left(Y_{K-1}^{t-1}, Y_{K,t+1}^{n}; Y_{K-1,t} \middle| X_K^n\right) \\ &\qquad + \sum_{t=1}^{n} I\left(X_K^n; Y_{K,t}\right) - \sum_{t=1}^{n} I\left(Y_{K,t+1}^{n}; Y_{K-1,t} \middle| X_{K-1}^n, X_K^n, Y_{K-1}^{t-1}\right) \\ &\quad = \sum_{t=1}^{n} I\left(X_{K-1}^n, Y_{K-1}^{t-1}, Y_{K,t+1}^{n}; Y_{K-1,t} \middle| X_K^n\right) + \sum_{t=1}^{n} I\left(X_K^n; Y_{K,t}\right) - \sum_{t=1}^{n} I\left(Y_{K,t+1}^{n}; Y_{K-1,t} \middle| X_{K-1}^n, X_K^n, Y_{K-1}^{t-1}\right) \\ &\quad = \sum_{t=1}^{n} I\left(X_{K-1}^n, Y_{K-1}^{t-1}; Y_{K-1,t} \middle| X_K^n\right) + \sum_{t=1}^{n} I\left(X_K^n; Y_{K,t}\right) \end{aligned} \tag{21}$$

Next, consider the following equality:

$$I(X_{K-2}^n; Y_{K-2}^n | X_{K-1}^n, X_K^n) = \sum_{t=1}^{n} I\left(X_{K-2}^n; Y_{K-2,t} \middle| X_{K-1}^n, X_K^n, Y_{K-2,t+1}^{n}\right) \tag{22}$$

Note that the style of applying the chain rule in (22) is similar to the second relation in (17); *it changes alternately among the mutual information functions in (16)*. Now consider the ensemble in (22) and the first ensemble on the right side of the last equality of (21); we can write:

$$\begin{aligned} &\sum_{t=1}^{n} I\left(X_{K-2}^n; Y_{K-2,t} \middle| X_{K-1}^n, X_K^n, Y_{K-2,t+1}^{n}\right) + \sum_{t=1}^{n} I\left(X_{K-1}^n, Y_{K-1}^{t-1}; Y_{K-1,t} \middle| X_K^n\right) \\ &\quad = \sum_{t=1}^{n} I\left(X_{K-2}^n, Y_{K-1}^{t-1}; Y_{K-2,t} \middle| X_{K-1}^n, X_K^n, Y_{K-2,t+1}^{n}\right) + \sum_{t=1}^{n} I\left(X_{K-1}^n, Y_{K-2,t+1}^{n}, Y_{K-1}^{t-1}; Y_{K-1,t} \middle| X_K^n\right) \\ &\qquad - \sum_{t=1}^{n} I\left(Y_{K-1}^{t-1}; Y_{K-2,t} \middle| X_{K-2}^n, X_{K-1}^n, X_K^n, Y_{K-2,t+1}^{n}\right) - \sum_{t=1}^{n} I\left(Y_{K-2,t+1}^{n}; Y_{K-1,t} \middle| X_{K-1}^n, X_K^n, Y_{K-1}^{t-1}\right) \\ &\quad = \sum_{t=1}^{n} I\left(X_{K-2}^n; Y_{K-2,t} \middle| X_{K-1}^n, X_K^n, Y_{K-1}^{t-1}, Y_{K-2,t+1}^{n}\right) + \sum_{t=1}^{n} I\left(Y_{K-1}^{t-1}; Y_{K-2,t} \middle| X_{K-1}^n, X_K^n, Y_{K-2,t+1}^{n}\right) \\ &\qquad + \sum_{t=1}^{n} I\left(Y_{K-2,t+1}^{n}, Y_{K-1}^{t-1}; Y_{K-1,t} \middle| X_{K-1}^n, X_K^n\right) + \sum_{t=1}^{n} I\left(X_{K-1}^n; Y_{K-1,t} \middle| X_K^n\right) \\ &\qquad - \sum_{t=1}^{n} I\left(Y_{K-1}^{t-1}; Y_{K-2,t} \middle| X_{K-2}^n, X_{K-1}^n, X_K^n, Y_{K-2,t+1}^{n}\right) - \sum_{t=1}^{n} I\left(Y_{K-2,t+1}^{n}; Y_{K-1,t} \middle| X_{K-1}^n, X_K^n, Y_{K-1}^{t-1}\right) \\ &\quad \overset{(a)}{=} \sum_{t=1}^{n} I\left(X_{K-2}^n; Y_{K-2,t} \middle| X_{K-1}^n, X_K^n, Y_{K-1}^{t-1}, Y_{K-2,t+1}^{n}\right) + \sum_{t=1}^{n} I\left(Y_{K-2,t+1}^{n}, Y_{K-1}^{t-1}; Y_{K-1,t} \middle| X_{K-1}^n, X_K^n\right) \\ &\qquad + \sum_{t=1}^{n} I\left(X_{K-1}^n; Y_{K-1,t} \middle| X_K^n\right) - \sum_{t=1}^{n} I\left(Y_{K-1}^{t-1}; Y_{K-2,t} \middle| X_{K-2}^n, X_{K-1}^n, X_K^n, Y_{K-2,t+1}^{n}\right) \end{aligned} \tag{23}$$

where equality (a) holds because, according to the Csiszar-Korner identity, the $2^{nd}$ and the $6^{th}$ ensembles on the left side of (a) are equal. Now consider the second ensemble on the right side of (a) in (87). We claim:

$$\sum_{t=1}^{n} I\left(Y_{K-2,t+1}^{n}, Y_{K-1}^{t-1}; Y_{K-1,t} \middle| X_{K-1}^n, X_K^n\right) \le \sum_{t=1}^{n} I\left(Y_{K-2,t+1}^{n}, Y_{K-1}^{t-1}; Y_{K-2,t} \middle| X_{K-1}^n, X_K^n\right) \tag{24}$$

To prove this inequality, first note that we have:

$$\begin{cases} \sum_{t=1}^{n} I\left(Y_{K-2,t+1}^{n}, Y_{K-1}^{t-1}; Y_{K-1,t} \middle| X_{K-1}^n, X_K^n\right) = \sum_{t=1}^{n} I\left(Y_{K-2,t+1}^{n}, Y_{K-1}^{t-1}; Y_{K-1,t} \middle| X_{K-1,t}, X_{K,t}, X_{K-1}^{n\backslash t}, X_K^{n\backslash t}\right) \\ \sum_{t=1}^{n} I\left(Y_{K-2,t+1}^{n}, Y_{K-1}^{t-1}; Y_{K-2,t} \middle| X_{K-1}^n, X_K^n\right) = \sum_{t=1}^{n} I\left(Y_{K-2,t+1}^{n}, Y_{K-1}^{t-1}; Y_{K-2,t} \middle| X_{K-1,t}, X_{K,t}, X_{K-1}^{n\backslash t}, X_K^{n\backslash t}\right) \end{cases} \tag{25}$$

By considering these equalities, (24) is derived from the condition (14) for $i = K-1, U \equiv \left(Y_{K-2,t+1}^{n}, Y_{K-1}^{t-1}\right)$ and $D \equiv \left(X_{K-1}^{n\backslash t}, X_{K}^{n\backslash t}\right)$. By substituting (24) in (23), we obtain:

$$
\begin{aligned}
&\sum_{t=1}^{n} I\left(X_{K-2}^{n}; Y_{K-2,t}\middle|X_{K-1}^{n}, X_{K}^{n}, Y_{K-2,t+1}^{n}\right) + \sum_{t=1}^{n} I\left(X_{K-1}^{n}, Y_{K-1}^{t-1}; Y_{K-1,t}\middle|X_{K}^{n}\right) \\
&\quad \leq \sum_{t=1}^{n} I\left(X_{K-2}^{n}; Y_{K-2,t}\middle|X_{K-1}^{n}, X_{K}^{n}, Y_{K-1}^{t-1}, Y_{K-2,t+1}^{n}\right) + \sum_{t=1}^{n} I\left(Y_{K-2,t+1}^{n}, Y_{K-1}^{t-1}; Y_{K-2,t}\middle|X_{K-1}^{n}, X_{K}^{n}\right) \\
&\qquad + \sum_{t=1}^{n} I\left(X_{K-1}^{n}; Y_{K-1,t}\middle|X_{K}^{n}\right) - \sum_{t=1}^{n} I\left(Y_{K-1}^{t-1}; Y_{K-2,t}\middle|X_{K-2}^{n}, X_{K-1}^{n}, X_{K}^{n}, Y_{K-2,t+1}^{n}\right) \\
&\quad = \sum_{t=1}^{n} I\left(X_{K-2}^{n}, Y_{K-1}^{t-1}, Y_{K-2,t+1}^{n}; Y_{K-2,t}\middle|X_{K-1}^{n}, X_{K}^{n}\right) + \sum_{t=1}^{n} I\left(X_{K-1}^{n}; Y_{K-1,t}\middle|X_{K}^{n}\right) \\
&\qquad - \sum_{t=1}^{n} I\left(Y_{K-1}^{t-1}; Y_{K-2,t}\middle|X_{K-2}^{n}, X_{K-1}^{n}, X_{K}^{n}, Y_{K-2,t+1}^{n}\right) \\
&\quad = \sum_{t=1}^{n} I\left(X_{K-2}^{n}, Y_{K-2,t+1}^{n}; Y_{K-2,t}\middle|X_{K-1}^{n}, X_{K}^{n}\right) + \sum_{t=1}^{n} I\left(X_{K-1}^{n}; Y_{K-1,t}\middle|X_{K}^{n}\right)
\end{aligned}
\tag{26}
$$

Therefore, by combining (21), (22) and (26), we have:

$$
\begin{aligned}
&I(X_{K-2}^{n}; Y_{K-2}^{n}|X_{K-1}^{n}, X_{K}^{n}) + I(X_{K-1}^{n}; Y_{K-1}^{n}|X_{K}^{n}) + I(X_{K}^{n}; Y_{K}^{n}) \\
&\quad \leq \sum_{t=1}^{n} I\left(X_{K-2}^{n}, Y_{K-2,t+1}^{n}; Y_{K-2,t}\middle|X_{K-1}^{n}, X_{K}^{n}\right) + \sum_{t=1}^{n} I\left(X_{K-1}^{n}; Y_{K-1,t}\middle|X_{K}^{n}\right) + \sum_{t=1}^{n} I\left(X_{K}^{n}; Y_{K,t}\right)
\end{aligned}
\tag{27}
$$

This procedure can be followed sequentially to manipulate other mutual information functions in (16). At last, we derive:

$$
I(X_{1}^{n}; Y_{1}^{n}|X_{2}^{n}, \dots, X_{K-1}^{n}, X_{K}^{n}) + \cdots + I(X_{K-1}^{n}; Y_{K-1}^{n}|X_{K}^{n}) + I(X_{K}^{n}; Y_{K}^{n}) \leq \Xi \tag{28}
$$

where:

- ✓ If $K$ is even, $\Xi$ is given by:

$$
\Xi = \sum_{t=1}^{n} I\left(X_{1}^{n}, Y_{1}^{t-1}; Y_{1,t}\middle|X_{2}^{n}, \dots, X_{K-1}^{n}, X_{K}^{n}\right) + \cdots + \sum_{t=1}^{n} I\left(X_{K-1}^{n}; Y_{K-1,t}\middle|X_{K}^{n}\right) + \sum_{t=1}^{n} I\left(X_{K}^{n}; Y_{K,t}\right) \tag{29}
$$

- ✓ If $K$ is odd, $\Xi$ is given by:

$$
\Xi = \sum_{t=1}^{n} I\left(X_{1}^{n}, Y_{1,t+1}^{n}; Y_{1,t}\middle|X_{2}^{n}, \dots, X_{K-1}^{n}, X_{K}^{n}\right) + \cdots + \sum_{t=1}^{n} I\left(X_{K-1}^{n}; Y_{K-1,t}\middle|X_{K}^{n}\right) + \sum_{t=1}^{n} I\left(X_{K}^{n}; Y_{K,t}\right) \tag{30}
$$

The expressions (29) and (30) both are in fact identical and equal to:

$$
\Xi = \sum_{t=1}^{n} I\left(X_{1}^{n}; Y_{1,t}\middle|X_{2}^{n}, \dots, X_{K-1}^{n}, X_{K}^{n}\right) + \cdots + \sum_{t=1}^{n} I\left(X_{K-1}^{n}; Y_{K-1,t}\middle|X_{K}^{n}\right) + \sum_{t=1}^{n} I\left(X_{K}^{n}; Y_{K,t}\right) \tag{31}
$$

This is because:

$$
\sum_{t=1}^{n} I\left(Y_{1}^{t-1}; Y_{1,t}\middle|X_{1}^{n}, X_{2}^{n}, \dots, X_{K-1}^{n}, X_{K}^{n}\right) = \sum_{t=1}^{n} I\left(Y_{1,t+1}^{n}; Y_{1,t}\middle|X_{1}^{n}, X_{2}^{n}, \dots, X_{K-1}^{n}, X_{K}^{n}\right) = 0 \tag{32}
$$

Moreover, since the network is memoryless, the following factorization is induced on the code:

$$
P(x_{1}^{n}, \dots, x_{K}^{n}, y_{1}^{n}, \dots, y_{K}^{n}) = \prod_{i=1}^{K} P(x_{i}^{n}) \prod_{t=1}^{n} \mathbb{P}\left(y_{1,t}, y_{2,t}, \dots, y_{K,t}\middle|x_{1,t}, x_{2,t}, \dots, x_{K,t}\right) \tag{33}
$$

Based on the factorization (33), one can easily show that for any $\Omega \subseteq \{1, \dots, K\}$ and any $i \in \{1, \dots, K\}$, the following Markov relations hold:

$$\{X_j^{n\setminus t}\}_{j\in\Omega} \to \{X_{j,t}\}_{j\in\Omega} \to Y_{i,t}, \qquad t = 1,\dots,n \tag{34}$$

Therefore, (31) can be simplified in the following single-letter form:

$$\Xi \triangleq \sum_{t=1}^{n} I\big(X_{1,t};Y_{1,t}\big|X_{2,t},\dots,X_{K-1,t},X_{K,t}\big) + \cdots + \sum_{t=1}^{n} I\big(X_{K-1,t};Y_{K-1,t}\big|X_{K,t}\big) + \sum_{t=1}^{n} I\big(X_{K,t};Y_{K,t}\big) \tag{35}$$

Finally, by applying a standard time-sharing argument, we derive the desired outer bound as given in (13). ■

In the next theorem, we show that the outer bound derived in Theorem 1 is indeed optimal for a class of ICs with a sequence of less noisy receivers.

***Theorem 2.*** *Consider the K-user IC given in Fig. 1. Assume that the channel satisfies the following less noisy conditions:*

$$I(U;Y_i|X_{i+1},X_{i+2},\dots,X_K) \leq \min\{I(U;Y_1|X_{i+1},X_{i+2},\dots,X_K), I(U;Y_2|X_{i+1},X_{i+2},\dots,X_K),\dots, I(U;Y_{i-1}|X_{i+1},X_{i+2},\dots,X_K)\} \tag{36}$$

*for all joint PDFs* $P_{UX_1X_2\dots X_i}P_{X_{i+1}}\dots P_{X_K}$, $i = 1,\dots,K$. *Then, the sum-rate capacity of the network is given by (13).*

*Proof of Theorem 2)* First note that according to Lemma 2, the conditions (36) can be extended as follows:

$$I(U;Y_i|X_{i+1},X_{i+2},\dots,X_K,D) \leq \min\{I(U;Y_1|X_{i+1},X_{i+2},\dots,X_K,D), I(U;Y_2|X_{i+1},X_{i+2},\dots,X_K,D),\dots, I(U;Y_{i-1}|X_{i+1},X_{i+2},\dots,X_K,D)\} \tag{37}$$

for all joint PDFs $P_{DUX_1X_2\dots X_K}$, $i = 1,\dots,K$. Now, for each $i$, if we set $D \equiv X_i$ in (37), we deduce that (37) implies (12). Therefore, if (36) holds, then the conditions of Theorem 1 are satisfied and thereby (13) constitutes an outer bound on the sum-rate capacity. It remains to prove that (13) is achievable as well. Consider the following simple successive decoding scheme. All messages are simply encoded at the transmitters similar to a multiple access channel (let the parameter $Q$ be a time-sharing variable). The receiver $Y_K$ decodes its own message and treats other signals as noise. The rate cost due to this step is:

$$I(X_K;Y_K|Q)$$

At the receiver $Y_{K-1}$, first the message corresponding to the receiver $Y_K$ is decoded. This step does not introduce any new rate cost. To see this consider (37) for $i = K$, $U \equiv X_K$, and $D \equiv Q$. We obtain:

$$I(X_K;Y_K|Q) \leq \min\{I(X_K;Y_1|Q), I(X_K;Y_2|Q),\dots, I(X_K;Y_{K-1}|Q)\} \leq I(X_K;Y_{K-1}|Q)$$

The receiver $Y_{K-1}$ next deocodes its own message. The rate cost due to this step is:

$$I(X_{K-1};Y_{K-1}|X_K,Q)$$

This decoding scheme is repeated at the other receivers similarly where the receiver $Y_i$ successively decodes all the messages corresponding to the receivers $Y_j$ with $i \leq j$. Considering the conditions (37), one can easily see that by this scheme a sum-rate equal to (13) is achieved. The proof is thus complete. ■

Consider the degraded interference network for which $X_1, X_2,\dots,X_K \to Y_1 \to Y_2 \to \cdots \to Y_{K_2}$ forms a Markov chain. For this network, all the less-noisy conditions in (36) are satisfied. Thus, Theorem 2 provides an alternative proof for a partial result of [18]. Note that if the matrix gain of the Gaussian network (2) is of rank one, it would be degraded and therefore its sum-rate capacity can be derived from Theorem 2.

In what follows, we present some interesting generalizations of the outer bound derived in Theorem 1. Accordingly, we obtain new bounds which can used to prove capacity results for many other scenarios.

To present the main result, we need to the following lemma.

***Lemma 5.*** *Consider the general interference channel in Fig. 1. Let* $Y_a$ *and* $Y_b$ *be two arbitrary receivers. Also, for any arbitrary* $\Omega_1, \Omega_2 \subseteq \{1, \dots, K\}$, *let* $\{X_l\}_{l\in\Omega_1}$ *and* $\{X_l\}_{l\in\Omega_2}$ *be two subset of input signals. Assume that the following condition holds:*

$$I\left(U, \{X_l\}_{l\in\Omega_1}; Y_a \middle| \{X_l\}_{l\in\Omega_2}\right) \leq I\left(U, \{X_l\}_{l\in\Omega_1}; Y_b \middle| \{X_l\}_{l\in\Omega_2}\right) \tag{38}$$

*for all joint PDFs* $P_{U,\{X_i\}_{i=1}^K - \{\{X_l\}_{l\in\Omega_2}\}} \prod_{X_i\in\{X_l\}_{l\in\Omega_2}} P_{X_i}$. *Then, given any code of length* $n$ *for the network, we have:*

$$I\left(\{X_l^n\}_{l\in\Omega_1}; Y_a^n \middle| \{X_l^n\}_{l\in\Omega_2}\right) \leq I\left(\{X_l^n\}_{l\in\Omega_1}; Y_b^n \middle| \{X_l^n\}_{l\in\Omega_2}\right) \tag{39}$$

*Proof of Lemma 5)* First note that the condition (38) can be extended to:

$$I\left(U, \{X_l\}_{l\in\Omega_1}; Y_a \middle| \{X_l\}_{l\in\Omega_2}, D\right) \leq I\left(U, \{X_l\}_{l\in\Omega_1}; Y_b \middle| \{X_l\}_{l\in\Omega_2}, D\right), \quad \text{for all joint PDFs } P_{DUX_1X_2\dots X_K} \tag{40}$$

This can be derived by following the same arguments as the proof of Lemma 2. Now, consider a code of length $n$ for the network. The proof indeed involves interesting computations including subtle applications of Csiszar-Korner identity. We have:

$$\begin{aligned}
&I\left(\{X_l^n\}_{l\in\Omega_1}; Y_a^n \middle| \{X_l^n\}_{l\in\Omega_2}\right) - I\left(\{X_l^n\}_{l\in\Omega_1}; Y_b^n \middle| \{X_l^n\}_{l\in\Omega_2}\right)\\
&= \textstyle\sum_{t=1}^n I\left(\{X_l^n\}_{l\in\Omega_1}; Y_{a,t} \middle| \{X_l^n\}_{l\in\Omega_2}, Y_a^{t-1}\right) - \sum_{t=1}^n I\left(\{X_l^n\}_{l\in\Omega_1}; Y_{b,t} \middle| \{X_l^n\}_{l\in\Omega_2}, Y_{b,t+1}^n\right)\\
&= \textstyle\sum_{t=1}^n I\left(\{X_l^n\}_{l\in\Omega_1}, Y_{b,t+1}^n; Y_{a,t} \middle| \{X_l^n\}_{l\in\Omega_2}, Y_a^{t-1}\right) - \sum_{t=1}^n I\left(\{X_l^n\}_{l\in\Omega_1}, Y_a^{t-1}; Y_{b,t} \middle| \{X_l^n\}_{l\in\Omega_2}, Y_{b,t+1}^n\right)\\
&\qquad - \textstyle\sum_{t=1}^n I\left(Y_{b,t+1}^n; Y_{a,t} \middle| \{X_l^n\}_{l\in\Omega_1}, \{X_l^n\}_{l\in\Omega_2}, Y_a^{t-1}\right) + \sum_{t=1}^n I\left(Y_a^{t-1}; Y_{b,t} \middle| \{X_l^n\}_{l\in\Omega_1}, \{X_l^n\}_{l\in\Omega_2}, Y_{b,t+1}^n\right)\\
&\overset{(a)}{=} \textstyle\sum_{t=1}^n I\left(\{X_l^n\}_{l\in\Omega_1}, Y_{b,t+1}^n; Y_{a,t} \middle| \{X_l^n\}_{l\in\Omega_2}, Y_a^{t-1}\right) - \sum_{t=1}^n I\left(\{X_l^n\}_{l\in\Omega_1}, Y_a^{t-1}; Y_{b,t} \middle| \{X_l^n\}_{l\in\Omega_2}, Y_{b,t+1}^n\right)\\
&= \textstyle\sum_{t=1}^n I\left(\{X_l^n\}_{l\in\Omega_1}; Y_{a,t} \middle| \{X_l^n\}_{l\in\Omega_2}, Y_a^{t-1}, Y_{b,t+1}^n\right) + \sum_{t=1}^n I\left(Y_{b,t+1}^n; Y_{a,t} \middle| \{X_l^n\}_{l\in\Omega_2}, Y_a^{t-1}\right)\\
&\qquad - \textstyle\sum_{t=1}^n I\left(\{X_l^n\}_{l\in\Omega_1}; Y_{b,t} \middle| \{X_l^n\}_{l\in\Omega_2}, Y_a^{t-1}, Y_{b,t+1}^n\right) - \sum_{t=1}^n I\left(Y_a^{t-1}; Y_{b,t} \middle| \{X_l^n\}_{l\in\Omega_2}, Y_{b,t+1}^n\right)\\
&\overset{(b)}{=} \textstyle\sum_{t=1}^n I\left(\{X_l^n\}_{l\in\Omega_1}; Y_{a,t} \middle| \{X_l^n\}_{l\in\Omega_2}, Y_a^{t-1}, Y_{b,t+1}^n\right) - \sum_{t=1}^n I\left(\{X_l^n\}_{l\in\Omega_1}; Y_{b,t} \middle| \{X_l^n\}_{l\in\Omega_2}, Y_a^{t-1}, Y_{b,t+1}^n\right)\\
&= \textstyle\sum_{t=1}^n I\left(\left\{X_l^{n\setminus t}\right\}_{l\in\Omega_1}, \left\{X_{l,t}\right\}_{l\in\Omega_1}; Y_{a,t} \middle| \left\{X_{l,t}\right\}_{l\in\Omega_2}, \left\{X_l^{n\setminus t}\right\}_{l\in\Omega_2}, Y_a^{t-1}, Y_{b,t+1}^n\right)\\
&\qquad - \textstyle\sum_{t=1}^n I\left(\left\{X_l^{n\setminus t}\right\}_{l\in\Omega_1}, \left\{X_{l,t}\right\}_{l\in\Omega_1}; Y_{b,t} \middle| \left\{X_{l,t}\right\}_{l\in\Omega_2}, \left\{X_l^{n\setminus t}\right\}_{l\in\Omega_2}, Y_a^{t-1}, Y_{b,t+1}^n\right)\\
&\overset{(c)}{\leq} 0
\end{aligned} \tag{41}$$

where equality (a) holds because, by Csiszar-Korner identity, the third and the fourth ensembles on the left side of (a) are equal; also, equality (b) holds because, again by Csiszar-Korner identity, the second and the fourth ensembles on the left side of (b) are equal; lastly, inequality (c) is derived from (40) by substituting $U \equiv \left\{X_l^{n\setminus t}\right\}_{l\in\Omega_1}$, and $D \equiv \left(\left\{X_l^{n\setminus t}\right\}_{l\in\Omega_2}, Y_a^{t-1}, Y_{b,t+1}^n\right)$. The proof is thus complete. ■

***Remark 1.*** Consider the condition (38). If $\Omega_1 = \{1, \dots, K\} - \Omega_2$, then the auxiliary random variable $U$ in (38) is dropped. In other words, it is reduced as follows:

$$I\left(\{X_l\}_{l\in\Omega_1}; Y_a \middle| \{X_l\}_{l\in\Omega_2}\right) \leq I\left(\{X_l\}_{l\in\Omega_1}; Y_b \middle| \{X_l\}_{l\in\Omega_2}\right) \tag{42}$$

for all joint PDFs $\prod_{i=1,\dots,K} P_{X_i}$. This is because $U \to \{X_1, X_2, \dots, X_K\} \to Y_a, Y_b$ forms a Markov chain.

Now, using Lemma 5, we derive the following generalization of Theorem 1.

***Theorem 3.*** *Consider the general K-user interference channel in Fig. 1. Let* $\lambda(.)$ *be a permutation of the elements of the set* $\{1, \dots, K\}$*. Let also* $i_1, i_2, \dots, i_\mu$ *be elements of the set* $\{1, \dots, K\}$ *with:*

$$1 \le i_1 < i_2 < \cdots < i_{\mu-1} < i_\mu = K, \tag{43}$$

*where* $\mu$ *is an arbitrary natural number less than or equal to* $K$*. Define:*

$$\mathbb{X}_{i_\theta} \triangleq \bigcup_{i_{\theta-1}+1 \le l \le i_\theta} \{X_{\lambda(l)}\}, \qquad \theta = 1, \dots, \mu, \qquad (i_0 \triangleq 0) \tag{44}$$

*Assume that the network transition probability function satisfies the following conditions:*

- ✓ *For* $\omega = 1, \dots, i_1 - 1 \implies$

$$I\left(\{X_{\lambda(l)}\}_{l\le\omega}; Y_{\lambda(\omega)} \middle| \{X_{\lambda(l)}\}_{\omega+1\le l}\right) \le I\left(\{X_{\lambda(l)}\}_{l\le\omega}; Y_{\lambda(\omega+1)} \middle| \{X_{\lambda(l)}\}_{\omega+1\le l}\right)$$
$$\text{for all joint PDFs} \quad P_{\{X_{\lambda(l)}\}_{l\le\omega}} \prod_{j\,\in\,\{\lambda(l)\}_{\omega+1\le l}} P_{X_j}, \tag{45}$$

- ✓ *For* $\theta = 1, \dots, \mu-1,\ \omega = 1, \dots, i_{\theta+1} - i_\theta - 1 \implies$

$$I\left(U, \{X_{\lambda(l)}\}_{i_\theta+1\le l\le i_\theta+\omega}; Y_{\lambda(i_\theta+\omega)} \middle| \{X_{\lambda(l)}\}_{i_\theta+\omega+1\le l}\right) \le I\left(U, \{X_{\lambda(l)}\}_{i_\theta+1\le l\le i_\theta+\omega}; Y_{\lambda(i_\theta+\omega+1)} \middle| \{X_{\lambda(l)}\}_{i_\theta+\omega+1\le l}\right)$$
$$\text{for all joint PDFs} \quad P_{U,\{X_{\lambda(l)}\}_{l\le i_\theta+\omega}} \prod_{j\,\in\,\{\lambda(l)\}_{i_\theta+\omega+1\le l}} P_{X_j}, \tag{46}$$

- ✓ *For* $\theta = 2, \dots, \mu \implies$

$$I\left(U; Y_{\lambda(i_\theta)} \middle| \cup_{\alpha=\theta}^{\mu} \mathbb{X}_{i_\alpha}\right) \le I\left(U; Y_{\lambda(i_{\theta-1})} \middle| \cup_{\alpha=\theta}^{\mu} \mathbb{X}_{i_\alpha}\right)$$
$$\text{for all joint PDFs} \quad P_{U,\{X_j\}_{j=1}^{K} - \cup_{\alpha=\theta}^{\mu} \mathbb{X}_{i_\alpha}} \prod_{X_j \in \cup_{\alpha=\theta}^{\mu} \mathbb{X}_{i_\alpha}} P_{X_j}, \tag{47}$$

*Then the sum-rate capacity is bounded-above by:*

$$\mathcal{C}_{sum} \le \max_{P_Q P_{X_1|Q} P_{X_2|Q} \cdots P_{X_K|Q}} \left( I\left(\mathbb{X}_{i_1}; Y_{\lambda(i_1)} \middle| \mathbb{X}_{i_2}, \dots, \mathbb{X}_{i_{\mu-1}}, \mathbb{X}_{i_\mu}, Q\right) + I\left(\mathbb{X}_{i_2}; Y_{\lambda(i_2)} \middle| \mathbb{X}_{i_3}, \dots, \mathbb{X}_{i_{\mu-1}}, \mathbb{X}_{i_\mu}, Q\right) + \cdots + I\left(\mathbb{X}_{i_\mu}; Y_{\lambda(i_\mu)} \middle| Q\right)\right) \tag{48}$$

*Proof of Theorem 3)* Consider a code of length $n$ for the network. Similar to (16), one can readily derive:

$$\begin{aligned}
n\mathcal{C}_{sum} \le\ & I\left(X_{\lambda(1)}^n; Y_{\lambda(1)}^n \middle| \{X_{\lambda(l)}^n\}_{2\le l}\right) + I\left(X_{\lambda(2)}^n; Y_{\lambda(2)}^n \middle| \{X_{\lambda(l)}^n\}_{3\le l}\right) + \cdots + I\left(X_{\lambda(i_1)}^n; Y_{\lambda(i_1)}^n \middle| \{X_{\lambda(l)}^n\}_{i_1+1\le l}\right) \\
& + I\left(X_{\lambda(i_1+1)}^n; Y_{\lambda(i_1+1)}^n \middle| \{X_{\lambda(l)}^n\}_{i_1+2\le l}\right) + \cdots + I\left(X_{\lambda(i_2)}^n; Y_{\lambda(i_2)}^n \middle| \{X_{\lambda(l)}^n\}_{i_2+1\le l}\right) \\
& + I\left(X_{\lambda(i_2+1)}^n; Y_{\lambda(i_2+1)}^n \middle| \{X_{\lambda(l)}^n\}_{i_2+2\le l}\right) + \cdots + I\left(X_{\lambda(i_3)}^n; Y_{\lambda(i_3)}^n \middle| \{X_{\lambda(l)}^n\}_{i_3+1\le l}\right) \\
& \qquad \vdots \\
& + I\left(X_{\lambda(i_{\mu-1}+1)}^n; Y_{\lambda(i_{\mu-1}+1)}^n \middle| \{X_{\lambda(l)}^n\}_{i_{\mu-1}+2\le l}\right) + \cdots + I\left(X_{\lambda(i_\mu)}^n; Y_{\lambda(i_\mu)}^n\right) \\
& + n\epsilon_n
\end{aligned} \tag{49}$$

where $\epsilon_n \to 0$ as $n \to 0$. Now consider the first row of (49). The conditions (45), according to Lemma 5 (see also Remark 1), imply that:

$$I\left(X_{\lambda(1)}^n; Y_{\lambda(1)}^n \middle| \left\{X_{\lambda(l)}^n\right\}_{2\le l}\right) + I\left(X_{\lambda(2)}^n; Y_{\lambda(2)}^n \middle| \left\{X_{\lambda(l)}^n\right\}_{3\le l}\right) + \cdots + I\left(X_{\lambda(i_1)}^n; Y_{\lambda(i_1)}^n \middle| \left\{X_{\lambda(l)}^n\right\}_{i_1+1\le l}\right)$$

$$\le I\left(X_{\lambda(1)}^n; Y_{\lambda(2)}^n \middle| \left\{X_{\lambda(l)}^n\right\}_{2\le l}\right) + I\left(X_{\lambda(2)}^n; Y_{\lambda(2)}^n \middle| \left\{X_{\lambda(l)}^n\right\}_{3\le l}\right) + I\left(X_{\lambda(3)}^n; Y_{\lambda(3)}^n \middle| \left\{X_{\lambda(l)}^n\right\}_{4\le l}\right) + \cdots + I\left(X_{\lambda(i_1)}^n; Y_{\lambda(i_1)}^n \middle| \left\{X_{\lambda(l)}^n\right\}_{i_1+1\le l}\right)$$

$$= I\left(X_{\lambda(1)}^n, X_{\lambda(2)}^n; Y_{\lambda(2)}^n \middle| \left\{X_{\lambda(l)}^n\right\}_{3\le l}\right) + I\left(X_{\lambda(3)}^n; Y_{\lambda(3)}^n \middle| \left\{X_{\lambda(l)}^n\right\}_{4\le l}\right) + \cdots + I\left(X_{\lambda(i_1)}^n; Y_{\lambda(i_1)}^n \middle| \left\{X_{\lambda(l)}^n\right\}_{i_1+1\le l}\right)$$

$$\le I\left(X_{\lambda(1)}^n, X_{\lambda(2)}^n; Y_{\lambda(3)}^n \middle| \left\{X_{\lambda(l)}^n\right\}_{3\le l}\right) + I\left(X_{\lambda(3)}^n; Y_{\lambda(3)}^n \middle| \left\{X_{\lambda(l)}^n\right\}_{4\le l}\right) + \cdots + I\left(X_{\lambda(i_1)}^n; Y_{\lambda(i_1)}^n \middle| \left\{X_{\lambda(l)}^n\right\}_{i_1+1\le l}\right)$$

$$= I\left(X_{\lambda(1)}^n, X_{\lambda(2)}^n, X_{\lambda(3)}^n; Y_{\lambda(3)}^n \middle| \left\{X_{\lambda(l)}^n\right\}_{4\le l}\right) + \cdots + I\left(X_{\lambda(i_1)}^n; Y_{\lambda(i_1)}^n \middle| \left\{X_{\lambda(l)}^n\right\}_{i_1+1\le l}\right)$$

$$\vdots$$

$$\le I\left(X_{\lambda(1)}^n, X_{\lambda(2)}^n, \ldots, X_{\lambda(i_1)}^n; Y_{\lambda(i_1)}^n \middle| \left\{X_{\lambda(l)}^n\right\}_{i_1+1\le l}\right) = I\left(\mathbb{X}_{i_1}^n; Y_{\lambda(i_1)}^n \middle| \mathbb{X}_{i_2}^n, \ldots, \mathbb{X}_{i_{\mu-1}}^n, \mathbb{X}_{i_\mu}^n\right) \tag{50}$$

Similarly, for the other rows in (49), according to Lemma 5, the conditions (46) imply that:

$$I\left(X_{\lambda(i_1+1)}^n; Y_{\lambda(i_1+1)}^n \middle| \left\{X_{\lambda(l)}^n\right\}_{i_1+2\le l}\right) + \cdots + I\left(X_{\lambda(i_2)}^n; Y_{\lambda(i_2)}^n \middle| \left\{X_{\lambda(l)}^n\right\}_{i_2+1\le l}\right) \le I\left(\mathbb{X}_{i_2}^n; Y_{\lambda(i_2)}^n \middle| \mathbb{X}_{i_3}^n, \ldots, \mathbb{X}_{i_{\mu-1}}^n, \mathbb{X}_{i_\mu}^n\right)$$

$$\vdots$$

$$I\left(X_{\lambda(i_{\mu-1}+1)}^n; Y_{\lambda(i_{\mu-1}+1)}^n \middle| \left\{X_{\lambda(l)}^n\right\}_{i_{\mu-1}+2\le l}\right) + \cdots + I\left(X_{\lambda(i_\mu)}^n; Y_{\lambda(i_\mu)}^n\right) \le I\left(\mathbb{X}_{i_\mu}^n; Y_{\lambda(i_\mu)}^n\right) \tag{51}$$

Thus, by substituting (50)-(51) in (49), we obtain:

$$n\mathcal{C}_{sum} \le I\left(\mathbb{X}_{i_1}^n; Y_{\lambda(i_1)}^n \middle| \mathbb{X}_{i_2}^n, \ldots, \mathbb{X}_{i_{\mu-1}}^n, \mathbb{X}_{i_\mu}^n\right) + I\left(\mathbb{X}_{i_2}^n; Y_{\lambda(i_2)}^n \middle| \mathbb{X}_{i_3}^n, \ldots, \mathbb{X}_{i_{\mu-1}}^n, \mathbb{X}_{i_\mu}^n\right) + \cdots + I\left(\mathbb{X}_{i_\mu}^n; Y_{\lambda(i_\mu)}^n\right) + n\epsilon_n \tag{52}$$

Finally, if the conditions (47) hold, by following the same lines as (16)-(35), one can derive the single-letter outer bound given in (48). The proof is thus complete. ■

***Remark 2.*** Let consider the two-user IC. By setting $\lambda(1) = 2$, $\lambda(2) = 1$, $\mu = 1$, $i_1 = 2$, in the conditions of Theorem 3, we obtain that if the following holds:

$$I(X_2; Y_2|X_1) \le I(X_2; Y_1|X_1) \quad \text{for all PDFs } P_{X_1}P_{X_2} \tag{53}$$

then, the sum-rate capacity is bounded by:

$$\mathcal{C}_{sum} \le \max_{P_Q P_{X_1|Q} P_{X_2|Q}} I(X_1, X_2; Y_1|Q) \tag{54}$$

Also, by setting $\lambda(1) = 1$, $\lambda(2) = 2$, $\mu = 2$, $i_1 = 1$, $i_2 = 2$, we obtain that if:

$$I(U; Y_2|X_2) \le I(U; Y_1|X_2) \quad \text{for all joint PDFs } P_{UX_2}P_{X_1} \tag{55}$$

then, the sum-rate capacity is bounded by:

$$\mathcal{C}_{sum} \leq \max_{P_Q P_{X_1|Q} P_{X_2|Q}} \big(I(X_1;Y_1|X_2,Q) + I(X_2;Y_2|Q)\big)$$

(56)

Therefore, one can deduce that if both the conditions (53) and (55) hold simultaneously, then the following is an outer bound on the sum-rate capacity:

$$\max_{P_Q P_{X_1|Q} P_{X_2|Q}} \min \begin{pmatrix} I(X_1,X_2;Y_1|Q) \\ I(X_1;Y_1|X_2,Q) + I(X_2;Y_2|Q) \end{pmatrix}$$

(57)

On the other hand, by a simple successive decoding scheme, one can achieve a sum-rate equal to (57). Thereby, (57) is the sum capacity of the channel under the conditions (53) and (55). These conditions in fact represent the *mixed interference regime* identified in [10, Th. 6] for the two-user IC. Thus, the result of [10, Th. 6] is indeed recovered by Theorem 3 here.

Next, we demonstrate how one can derive new sum-rate capacity results for many different interference networks using the outer bound established in Theorem 3. Specifically, we consider the three-user IC. The Gaussian channel is formulated as:

$$\begin{cases} Y_1 = X_1 + a_{12}X_2 + a_{13}X_3 + Z_1 \\ Y_2 = a_{21}X_1 + X_2 + a_{23}X_3 + Z_2 \\ Y_3 = a_{31}X_1 + a_{32}X_2 + X_3 + Z_3 \end{cases}$$

(58)

where $Z_1$, $Z_2$, and $Z_3$ are zero-mean unit-variance Gaussian random variables and the inputs are subject to power constraints $\mathbb{E}[X_i^2] \leq P_i, i = 1,2,3$.

First let us present the sum-rate achievable by the successive decoding scheme. It is given below:

$$\max_{P_Q P_{X_1|Q} P_{X_2|Q} P_{X_3|Q}} \min \begin{Bmatrix} I(X_1;Y_1|X_2,X_3,Q) + I(X_2:Y_2|X_3,Q) + I(X_3;Y_3|Q), \\ I(X_1;Y_1|X_2,X_3,Q) + I(X_2,X_3:Y_2|Q), \\ I(X_1;Y_1|X_2,X_3,Q) + I(X_2:Y_2|X_3,Q) + I(X_3;Y_1|Q), \\ I(X_1,X_2;Y_1|X_3,Q) + I(X_3;Y_3|Q), \\ I(X_1,X_2;Y_1|X_3,Q) + I(X_3;Y_2|Q), \\ I(X_1,X_2,X_3;Y_1|Q) \end{Bmatrix}$$

(59)

To achieve this sum-rate:

- ✓ The receiver $Y_3$ decodes only its corresponding signal $X_3$.
- ✓ The receiver $Y_2$ decodes the signal $X_3$ first and then decodes its corresponding signal $X_2$.
- ✓ The receiver $Y_1$ decodes the signal $X_3$ first, then the signal $X_2$, and lastly its corresponding signal $X_1$.

As we see from (59), the sum-rate expression due to this achievability scheme is described by six constraints. We intend to explore conditions under which this achievable sum-rate is optimal. Note that some of the constraints in (59) do not have a structure similar to the expression of the outer bound (48); for example, the third and the fifth ones. Therefore, we need to impose appropriate conditions on the network probability function so that such constraints can be relaxed. Let the network satisfies the following conditions:

$$\begin{cases} I(X_2:Y_2|X_3,Q) \geq I(X_2:Y_1|X_3,Q) \\ I(X_3;Y_2|Q) \geq I(X_3;Y_1|Q) \end{cases}$$

(60)

for all joint PDFs which are a solution to the following maximization:

$$\max_{P_Q P_{X_1|Q} P_{X_2|Q} P_{X_3|Q}} \min \begin{pmatrix} I(X_1,X_2;Y_1|X_3,Q) + I(X_3;Y_3|Q), \\ I(X_1,X_2,X_3;Y_1|Q) \end{pmatrix}$$

(61)

In this case, the achievable sum-rate (59) is reduced to (61). In other words, the conditions (60) enable us to relax those constraints of (59) which are not given in (61). Now consider the outer bound (48) specialized for the three-user IC. By setting $\lambda(1) = 2$, $\lambda(2) = 1$, $\lambda(3) = 3$, $\mu = 2$, $i_1 = 2$, and $i_2 = 3$, we obtain that if the following conditions hold:

$$\begin{cases} I(X_2;Y_2|X_1,X_3) \leq I(X_2;Y_1|X_1,X_3) & \text{for all joint PDFs} \quad P_{X_1}P_{X_2}P_{X_3} \\ I(U;Y_3|X_3) \leq I(U;Y_1|X_3) & \text{for all joint PDFs} \quad P_{UX_1X_2}P_{X_3} \end{cases} \tag{62}$$

then, the sum-rate capacity is bounded above by:

$$\mathcal{C}_{sum} \leq \max_{P_Q P_{X_1|Q} P_{X_2|Q} P_{X_3|Q}} \big(I(X_1,X_2;Y_1|X_3,Q) + I(X_3;Y_3|Q)\big) \tag{63}$$

Also, by setting $\lambda(1) = 3, \lambda(2) = 2, \lambda(3) = 1$, $\mu = 1$, and $i_1 = 3$, we obtain that if the following conditions hold:

$$\begin{cases} I(X_3;Y_3|X_1,X_2) \leq I(X_3;Y_2|X_1,X_2) & \text{for all joint PDFs} \quad P_{X_1}P_{X_2}P_{X_3} \\ I(X_2,X_3;Y_2|X_1) \leq I(X_2,X_3;Y_1|X_1) & \text{for all joint PDFs} \quad P_{X_1}P_{X_2X_3} \end{cases} \tag{64}$$

then, the sum-rate capacity is bounded above as:

$$\mathcal{C}_{sum} \leq \max_{P_Q P_{X_1|Q} P_{X_2|Q} P_{X_3|Q}} \big(I(X_1,X_2,X_3;Y_1|Q)\big) \tag{65}$$

Thus, if the collection of the conditions (60), (62) and (64) are satisfied, the sum-rate capacity of the network is given by (61). This capacity result could not be obtained using the outer bound (13) derived in Theorem 1. Let us now consider the Gaussian channel given in (58). Lemmas 3 and 4 imply that if the channel gains satisfy:

$$\begin{cases} |a_{12}| \geq 1, & |a_{31}| \leq 1, & |a_{23}| \geq 1 \\ a_{31} = \dfrac{a_{32}}{a_{12}}, & a_{12} = \dfrac{a_{13}}{a_{23}} & \end{cases}, \tag{66}$$

then (62) and (64) are also satisfied. Note that, according to Corollary 1, the second inequality of (64) implies the first inequality of (62). For a Gaussian channel satisfying (66), using the entropy power inequality, one can prove that Gaussian input distributions without time-sharing ($Q \equiv \emptyset$) is the solution to the maximization (61). Hence, if the inequalities (60) hold for Gaussian distributions, then the sum-rate capacity is given by (61). Considering (66), it is readily derived that both inequalities (60) hold for Gaussian distributions provided that:

$$P_1 + 1 \geq a_{12}^2(a_{21}^2 P_1 + 1) \tag{67}$$

Thus, for a three-user Gaussian IC (58) with the conditions (66)-(67), the sum-rate capacity is given as follows:

$$\min \begin{pmatrix} \psi(P_1 + a_{12}^2 P_2) + \psi\left(\dfrac{P_3}{a_{31}^2 P_1 + a_{32}^2 P_2 + 1}\right) \\ \psi(P_1 + a_{12}^2 P_2 + a_{13}^2 P_3) \end{pmatrix} \tag{68}$$

where $\psi(x) := \frac{1}{2}\log(1 + x)$. This sum-rate capacity is achieved by successive decoding scheme.

It should be remarked that by applying more efficient coding strategies such as joint decoding scheme (or combination of both successive and joint decoding schemes), one can achieve further capacity results for the ICs. Indeed, our approach could be followed to obtain sum-rate capacity for many other network topologies.

A second generalization of the result of Theorem 1 is given below.

***Theorem 4.*** *Consider the general K-user IC shown in Fig. 1. Let* $\boldsymbol{Y}_{G(1)}, \boldsymbol{Y}_{G(2)}, \dots, \boldsymbol{Y}_{G(\mu)}$ *be nonempty subsets of the set of the outputs* $\{Y_1, \dots, Y_K\}$ *so that:*

$$\begin{cases}\boldsymbol{Y}_{G(i_1)} \cap \boldsymbol{Y}_{G(i_2)} \neq \emptyset \\ \bigcup_{i=1}^{\mu} \boldsymbol{Y}_{G(i)} = \{Y_1, \dots, Y_K\}\end{cases} \tag{69}$$

*i.e., the collection* $\boldsymbol{Y}_{G(1)}, \boldsymbol{Y}_{G(2)}, \dots, \boldsymbol{Y}_{G(\mu)}$ *constitutes a nonempty partitioning for* $\{Y_1, \dots, Y_K\}$. *Define:*

$$\mathbb{X}_{G(i)} \triangleq \bigcup_{l:\, Y_l \in \boldsymbol{Y}_{G(i)}} \{X_l\}, \qquad i = 1, \dots, \mu \tag{70}$$

*Assume that the network transition probability function satisfies the following conditions:*

$$I\big(U; \boldsymbol{Y}_{G(i)} \big| \mathbb{X}_{G(i)}, \mathbb{X}_{G(i+1)}, \dots, \mathbb{X}_{G(\mu)}\big) \leq I\big(U; \boldsymbol{Y}_{G(i-1)} \big| \mathbb{X}_{G(i)}, \mathbb{X}_{G(i+1)}, \dots, \mathbb{X}_{G(\mu)}\big),$$
$$\text{for all joint PDFs} \quad P_{U,\{X_j\}_{j=1}^{K} - \cup_{\alpha=i}^{\mu} \mathbb{X}_{G(\alpha)}} \prod_{X_j \in \cup_{\alpha=i}^{\mu} \mathbb{X}_{G(\alpha)}} P_{X_j}, \qquad i = 2, \dots, \mu \tag{71}$$

*Then the sum-rate capacity of the network is bounded above as:*

$$\mathcal{C}_{sum} \leq \max_{P_Q P_{X_1|Q} P_{X_2|Q} \dots P_{X_K|Q}} \Big( I\big(\mathbb{X}_{G(1)}; \boldsymbol{Y}_{G(1)} \big| \mathbb{X}_{G(2)}, \dots, \mathbb{X}_{G(\mu-1)}, \mathbb{X}_{G(\mu)}, Q\big) + \dots + I\big(\mathbb{X}_{G(\mu-1)}; \boldsymbol{Y}_{G(\mu-1)} \big| \mathbb{X}_{G(\mu)}, Q\big) + I\big(\mathbb{X}_{G(\mu)}; \boldsymbol{Y}_{G(\mu)} \big| Q\big) \Big) \tag{72}$$

*Proof of Theorem 4)* Consider a code of length $n$ for the network with the rates $R_1, R_2, \dots, R_K$ for the users $X_1, X_2, \dots, X_K$, respectively. Given the definition (70), using Fano's inequality, one can derive:

$$\begin{aligned} n \sum_{l: Y_l \in \boldsymbol{Y}_{G(i)}} R_l &\leq I\big(\mathbb{X}_{G(i)}^n; \boldsymbol{Y}_{G(i)}^n\big) + n\epsilon_{G(i),n} \\ &\overset{(a)}{\leq} I\big(\mathbb{X}_{G(i)}^n; \boldsymbol{Y}_{G(i)}^n \big| \mathbb{X}_{G(i+1)}^n \cup \dots \cup \mathbb{X}_{G(\mu)}^n\big) + n\epsilon_{G(i),n} \end{aligned} \tag{73}$$

where $\epsilon_{G(i),n} \to 0$ as $n \to \infty$. Note that inequality (a) in (73) holds because the signals of the set $\mathbb{X}_{G(i)}^n$ are independent of those in $\mathbb{X}_{G(i+1)}^n \cup \dots \cup \mathbb{X}_{G(\mu)}^n$. By adding the two sides of (73) for $i = 1, \dots, \mu$, we obtain:

$$\begin{aligned} n\mathcal{C}_{sum} &= n \left( \sum_{i=1}^{\mu} \sum_{l: Y_l \in \boldsymbol{Y}_{G(i)}} R_l \right) \\ &\leq I\big(\mathbb{X}_{G(1)}^n; \boldsymbol{Y}_{G(1)}^n \big| \mathbb{X}_{G(2)}^n, \dots, \mathbb{X}_{G(\mu-1)}^n, \mathbb{X}_{G(\mu)}^n\big) + \dots + I\big(\mathbb{X}_{G(\mu-1)}^n; \boldsymbol{Y}_{G(\mu-1)}^n \big| \mathbb{X}_{G(\mu)}^n\big) + I\big(\mathbb{X}_{G(\mu)}^n; \boldsymbol{Y}_{G(\mu)}^n\big) + n\epsilon_n \end{aligned} \tag{74}$$

where $\epsilon_n \to 0$ as $n \to \infty$. Now, if the conditions (71) hold, by following the same lines as (16)-(35), one can derive the single-letter outer bound given in (72). The proof is thus complete. ■

***Remark 3.*** It is clear that by setting $\mu = K$ and $\boldsymbol{Y}_{G(i)} = \{Y_i\},\ i = 1, \dots, K$, the conditions (71) and the outer bound (72) are reduced to (12) and (13), respectively. In fact, the generalized bound of Theorem 4 could be simply deduced by considering a virtual interference network with the outputs $\boldsymbol{Y}_{G(1)}, \boldsymbol{Y}_{G(2)}, \dots, \boldsymbol{Y}_{G(\mu)}$ and the corresponding inputs $\mathbb{X}_{G(1)}, \mathbb{X}_{G(2)}, \dots, \mathbb{X}_{G(\mu)}$ and then applying the result of Theorem 1.

The outer bound given in Theorem 4 can be used to prove explicit sum-rate capacity results which are not necessarily derived from the bounds of Theorems 1 and 3. We conclude the paper by providing an example in this regard. Consider the K-user *many-to-one interference channel*. This is a special class of the K-user interference channel where only one receiver experiences interference. In this case, the channel transition probability function is factorized in the following form:

$$\mathbb{P}_{Y_1\dots Y_K|X_1\dots X_K}=\mathbb{P}_{Y_1|X_1}\mathbb{P}_{Y_2|X_2}\mathbb{P}_{Y_3|X_3}\dots\mathbb{P}_{Y_{K-1}|X_{K-1}}\mathbb{P}_{Y_K|X_1X_2X_3\dots X_K} \tag{75}$$

Define $\boldsymbol{Y}_{G(1)},\boldsymbol{Y}_{G(2)}$ as follows:

$$\boldsymbol{Y}_{G(1)}\triangleq\{Y_1,Y_2,\dots,Y_{K-1}\},\qquad \boldsymbol{Y}_{G(2)}\triangleq\{Y_K\} \tag{76}$$

Now by letting $\mu=2$ and substituting (76) in Theorem 4, we obtain that if the following condition holds:

$$I(U;Y_K|X_K)\leq I(U;Y_1,Y_2,\dots,Y_{K-1}|X_K),\quad \text{for all joint PDFs}\quad P_{UX_1X_2\dots X_{K-1}}P_{X_K}, \tag{77}$$

then the sum-rate capacity is bounded above as:

$$\begin{aligned}\mathcal{C}_{sum}&\leq\max_{P_QP_{X_1|Q}P_{X_2|Q}\dots P_{X_K|Q}}\big(I(X_1,X_2,\dots,X_{K-1};Y_1,Y_2,\dots,Y_{K-1}|X_K,Q)+I(X_K;Y_K|Q)\big)\\&\overset{(a)}{=}\max_{P_QP_{X_1|Q}P_{X_2|Q}\dots P_{X_K|Q}}\big(I(X_1;Y_1|Q)+I(X_2;Y_2|Q)+\dots+I(X_K;Y_K|Q)\big)\end{aligned} \tag{78}$$

where equality (a) is due to the factorization (75). Now, one can achieve (78) by a simple treating interference as noise strategy, i.e., the outer bound is in fact optimal. Thus, the sum-rate capacity of the many-to-one interference channel, if the less noisy condition (77) holds, is given by:

$$\mathcal{C}_{sum}=\max_{P_QP_{X_1|Q}P_{X_2|Q}\dots P_{X_K|Q}}\big(I(X_1;Y_1|Q)+I(X_2;Y_2|Q)+\dots+I(X_K;Y_K|Q)\big) \tag{79}$$

Similarly, many other scenarios can be identified for which the outer bound derived in Theorem 4 yields the exact sum-rate capacity.

# CONCLUSION

In this paper, we presented novel techniques to derive single-letter outer bounds on the sum-rate capacity of multi-user interference channels. Our approach requires subtle sequential applications of Csiszar-Korner identity. We demonstrated that our bounds are efficient to prove several new capacity results for various networks under specific conditions. Our new capacity results hold for both discrete and Gaussian channels.

## Appendix A

Proof of Lemma 1

First we show that (3) implies the following inequality:

$$I\big(X_1,\dots,X_{\mu_1};Y_1\big|X_{\mu_1+1},\dots,X_{\mu_1+\mu_2},W\big)\le I\big(X_1,\dots,X_{\mu_1};Y_2\big|X_{\mu_1+1},\dots,X_{\mu_1+\mu_2},W\big) \tag{80}$$

for all PDFs $P_{WX_1\dots X_{\mu_1}X_{\mu_1+1}\dots X_{\mu_1+\mu_2}}\big(w,x_1,\dots,x_{\mu_1},x_{\mu_1+1},\dots,x_{\mu_1+\mu_2}\big)$ with:

$$P_{WX_1\dots X_{\mu_1}X_{\mu_1+1}\dots X_{\mu_1+\mu_2}}=P_WP_{X_1\dots X_{\mu_1}|W}P_{X_{\mu_1+1}|W}P_{X_{\mu_1+2}|W}\dots P_{X_{\mu_1+\mu_2}|W} \tag{81}$$

where $W\to X_1,\dots,X_{\mu_1},X_{\mu_1+1},\dots,X_{\mu_1+\mu_2}\to Y_1,Y_2$ forms a Markov chain. To prove this inequality, one can write:

$$\begin{aligned}
&I\big(X_1,\dots,X_{\mu_1};Y_1\big|X_{\mu_1+1},\dots,X_{\mu_1+\mu_2},W\big)\\
&\quad=\sum_w P_W(w)I\big(X_1,\dots,X_{\mu_1};Y_1\big|X_{\mu_1+1},\dots,X_{\mu_1+\mu_2},w\big)\\
&\quad=\sum_w P_W(w)I\big(X_1,\dots,X_{\mu_1};Y_1\big|X_{\mu_1+1},\dots,X_{\mu_1+\mu_2}\big)_{\langle P_{X_1\dots X_{\mu_1}|w}\times P_{X_{\mu_1+1}|w}\times P_{X_{\mu_1+2}|w}\times\dots\times P_{X_{\mu_1+\mu_2}|w}\rangle}\\
&\quad\overset{(a)}{\le}\sum_w P_W(w)I\big(X_1,\dots,X_{\mu_1};Y_2\big|X_{\mu_1+1},\dots,X_{\mu_1+\mu_2}\big)_{\langle P_{X_1\dots X_{\mu_1}|w}\times P_{X_{\mu_1+1}|w}\times P_{X_{\mu_1+2}|w}\times\dots\times P_{X_{\mu_1+\mu_2}|w}\rangle}\\
&\quad=\sum_w P_W(w)I\big(X_1,\dots,X_{\mu_1};Y_2\big|X_{\mu_1+1},\dots,X_{\mu_1+\mu_2},w\big)\\
&\quad=I\big(X_1,\dots,X_{\mu_1};Y_2\big|X_{\mu_1+1},\dots,X_{\mu_1+\mu_2},W\big)
\end{aligned} \tag{82}$$

where the notation $I(A;B|C)_{\langle P(.)\rangle}$ indicates that the mutual information function $I(A;B|C)$ is evaluated by the distribution $P(.)$. Note that for any given $w$, the function $P_{X_1\dots X_{\mu_1}|w}\times P_{X_{\mu_1+1}|w}\times P_{X_{\mu_1+2}|w}\times\dots\times P_{X_{\mu_1+\mu_2}|w}$ is a probability distribution defined over the set $\mathcal{X}_1\times\dots\times\mathcal{X}_{\mu_1}\times\mathcal{X}_{\mu_1+1}\times\dots\times\mathcal{X}_{\mu_1+\mu_2}$ with the factorization (4). The inequality (a) is due to (3).

Now, having at hand the inequality (80), one can substitute $W\equiv\big(D,X_{\mu_1+1},X_{\mu_1+2},\dots,X_{\mu_1+\mu_2}\big)$ with an arbitrary joint distribution[1] on the set $\mathcal{D}\times\mathcal{X}_{\mu_1+1}\times\dots\times\mathcal{X}_{\mu_1+\mu_2}$. By this substitution, we derive that (5) holds for all joint PDFs $P_{DX_{\mu_1+1}\dots X_{\mu_1+\mu_2}}P_{X_1\dots X_{\mu_1}|DX_{\mu_1+1}\dots X_{\mu_1+\mu_2}}$. The proof is thus complete. ■

## Appendix B

Proof of Lemma 2

The proof is rather similar to Lemma 1. First, note that (7) implies the following inequality:

$$I\big(U;Y_1\big|X_{\mu_1+1},\dots,X_{\mu_1+\mu_2},W\big)\le I\big(U;Y_2\big|X_{\mu_1+1},\dots,X_{\mu_1+\mu_2},W\big) \tag{83}$$

for all PDFs $P_{WUX_1\dots X_{\mu_1}X_{\mu_1+1}\dots X_{\mu_1+\mu_2}}\big(w,u,x_1,\dots,x_{\mu_1},x_{\mu_1+1},\dots,x_{\mu_1+\mu_2}\big)$ with:

$$P_{WUX_1\dots X_{\mu_1}X_{\mu_1+1}\dots X_{\mu_1+\mu_2}}=P_WP_{UX_1\dots X_{\mu_1}|W}P_{X_{\mu_1+1}|W}P_{X_{\mu_1+2}|W}\dots P_{X_{\mu_1+\mu_2}|W} \tag{84}$$

[1] We have this liberty because $P_W(w)$ in (81) is arbitrary.

where $W, U \to X_1, \dots, X_{\mu_1}, X_{\mu_1+1}, \dots, X_{\mu_1+\mu_2} \to Y_1, Y_2$ forms a Markov chain. This can be proved by following the same lines as (82). Now, having at hand the inequality (83), one can substitute $W \equiv \left(X_{\mu_1+1}, X_{\mu_1+2}, \dots, X_{\mu_1+\mu_2}, D\right)$ with an arbitrary joint distribution on the set $\mathcal{X}_{\mu_1+1} \times \dots \times \mathcal{X}_{\mu_1+\mu_2} \times \mathcal{D}$ . By this substitution, we obtain that the inequality (9) holds for all joint PDFs $P_{X_{\mu_1+1}\dots X_{\mu_1+\mu_2}D} P_{UX_1\dots X_{\mu_1}|X_{\mu_1+1}\dots X_{\mu_1+\mu_2}D}$. The proof is complete. ■

## Appendix C

Proof of Lemma 3

First note that if $D$ is independent of $(Z_1, Z_2)$, then $D \to X_1, \dots, X_{\mu_1}, X_{\mu_1+1}, \dots, X_{\mu_1+\mu_2} \to Y_1, Y_2$ forms a Markov chain. It is sufficient to prove that (3) holds. Now define:

$$\tilde{Y}_1 \triangleq \alpha Y_2 + \left(\alpha b_{\mu_1+1} - a_{\mu_1+1}\right) X_{\mu_1+1} + \left(\alpha b_{\mu_1+2} - a_{\mu_1+2}\right) X_{\mu_1+2} + \dots + \left(\alpha b_{\mu_1+\mu_2} - a_{\mu_1+\mu_2}\right) X_{\mu_1+\mu_2} + \sqrt{1-\alpha^2}\tilde{Z}_1 \tag{85}$$

where $\tilde{Z}_1$ is a Gaussian random variable with zero mean and unit variance which is independent of $(Z_1, Z_2)$. By considering (10), it is readily derived that $\tilde{Y}_1$ is statistically equivalent to $Y_1$ in the sense of:

$$\mathbb{P}\left(\tilde{y}_1 \middle| x_1, \dots, x_{\mu_1}, x_{\mu_1+1}, \dots, x_{\mu_1+\mu_2}\right) \approx \mathbb{P}\left(y_1 \middle| x_1, \dots, x_{\mu_1}, x_{\mu_1+1}, \dots, x_{\mu_1+\mu_2}\right) \tag{86}$$

Therefore, for all input distributions, we have:

$$\begin{aligned}
I\left(X_1, \dots, X_{\mu_1}; Y_1 \middle| X_{\mu_1+1}, \dots, X_{\mu_1+\mu_2}\right) &= I\left(X_1, \dots, X_{\mu_1}; \tilde{Y}_1 \middle| X_{\mu_1+1}, \dots, X_{\mu_1+\mu_2}\right) \\
&\leq I\left(X_1, \dots, X_{\mu_1}; \tilde{Y}_1, Y_2 \middle| X_{\mu_1+1}, \dots, X_{\mu_1+\mu_2}\right) \\
&\overset{(a)}{=} I\left(X_1, \dots, X_{\mu_1}; Y_2 \middle| X_{\mu_1+1}, \dots, X_{\mu_1+\mu_2}\right) + I\left(X_1, \dots, X_{\mu_1}; \tilde{Y}_1 \middle| Y_2, X_{\mu_1+1}, \dots, X_{\mu_1+\mu_2}\right) \\
&= I\left(X_1, \dots, X_{\mu_1}; Y_2 \middle| X_{\mu_1+1}, \dots, X_{\mu_1+\mu_2}\right)
\end{aligned} \tag{87}$$

where (a) holds because $X_1, \dots, X_{\mu_1} \to Y_2, X_{\mu_1+1}, \dots, X_{\mu_1+\mu_2} \to \tilde{Y}_1$ forms a Markov chain (this is clear from (85)). The proof is thus complete. ■